\documentclass[aps,pra,floatfix,twocolumn,superscriptaddress,nofootinbib]{revtex4-1}
\usepackage{times}
\usepackage{amsmath,amssymb,verbatim,float,mathtools}
\usepackage[english]{babel}
\usepackage{hyperref}
\usepackage{txfonts}
\usepackage{enumitem}
\hypersetup{
     colorlinks   = true,
     citecolor    = blue,
    linkcolor=blue,
    urlcolor=blue
}

\begin{document}
\title{Size effects in superconducting thin films coupled to a substrate}
\author{Aurelio Romero-Berm\'udez }
\address{University of Cambridge, Cavendish Laboratory, JJ Thomson Avenue, Cambridge, CB3 0HE, United Knigdom}
\author{Antonio M. Garc\'{\i}a-Garc\'{\i}a}
\address{University of Cambridge, Cavendish Laboratory, JJ Thomson Avenue, Cambridge, CB3 0HE, United Knigdom}
\address{CFIF, Instituto Superior T\'ecnico, Universidade de Lisboa, Avenida Rovisco Pais, 1049-001 Lisboa, Portugal}{

\begin{abstract}
Recent experimental advances in surface science have made it possible to track the evolution of superconductivity in films as the thickness enters the nanoscale region where it is expected that the substrate plays an important role. Here, we put forward a mean-field, analytically tractable, model that describes size effects in ultrathin films coupled to the substrate. We restrict our study to one-band, crystalline, weakly coupled superconductors with no impurities. The thin-film substrate/vacuum interfaces are described by a simple asymmetric potential well and a finite quasiparticle lifetime. Boundary conditions are chosen to comply with the charge neutrality condition. This model provides a fair description of experimental results in ultrathin lead films: on average, the superconducting gap decreases with thickness and it is always below the bulk value. Clear oscillations, remnants of the shape resonances, are still observed for intermediate thicknesses. For materials with a weaker electron-phonon coupling and negligible disorder, a modest enhancement of superconductivity seems to be feasible. The relaxation of the charge neutrality condition, which is in principle justified in complex oxide heterostructures and other materials, would lead to a much stronger enhancement of superconductivity by size effects.
\end{abstract}
\maketitle

\section{Introduction}
Research on superconducting thin films has a long tradition in condensed matter physics. In the early 1960s, theoretical mean-field models \cite{Thompson1963} predicted oscillations of the superconducting gap and the critical temperature for nanosize film thickness with peaks that greatly exceeded the bulk limit. This nonmonotonic size dependence, usually referred to as shape resonances, has a simple origin. As thickness increases from the two-dimensional limit, new states become eventually available with the quantum numbers of an infinite well of size the thickness of the film. This additional subband enhances superconductivity as the spectral density is proportional to the dimensionless electron-phonon coupling constant. After the first peak, for larger thicknesses, the spectral density decreases until a new subband becomes available and a new peak occurs in the critical temperature.

Initial experimental results in granular thin films of Al \cite{abeles}  and other materials \cite{Kresin}  also reported a substantial enhancement of the critical temperature with respect to the bulk limit. However, granular materials are intrinsically disordered and impurities suppress shape resonances so a direct relation between theoretical and experimental results was hard to establish.

It was later realized \cite{Allen1975,Yu1976}  that no enhancement is observed in more realistic theoretical models that impose charge neutrality at the interfaces. More refined experiments with smoother films and a better experimental control \cite{goldman} observed no enhancement of superconductivity but rather a transition at a temperature lower than the bulk mean-field theory prediction.

Recent progresses in nanotechnology and surface science, in particular epitaxial deposition and scanning tunneling microscopy/spectroscopy (STM), have dramatically improved the experimental control in low dimensions, which has led to many exciting results \cite{Qin2009,Zhang2010,Brun2009,Bose2010}. For instance, experiments on ultrathin Pb films with thicknesses ranging from a single to a few atomic monolayers \cite{Guo2004,Qin2009,Zhang2010}  found that superconductivity is still present although weaker than in the bulk limit. Oscillations of the superconducting gap and the critical temperature, below the bulk value, for intermediate thickness, were also reported. Theoretical models proposed to described these results \cite{Chen2006,Shanenko2007} had free parameters and did not include important features such as the role of the substrate, the finite lifetime of quasiparticles, or an adequate description of the interface. As the thickness decreases, we expect that these features become increasingly important. More detailed first-principles calculations \cite{leadfirst} of the interface in the ultrathin limit do not address superconductivity explicitly. Strikingly, experimental results in oxide interfaces \cite{Reyren2007}, and even single- layered iron-based superconductors \cite{Liu2012}, exhibit, in some cases, an enhancement of the critical temperature with respect to the bulk limit. The theoretical reasons of this behavior are not yet well understood.

Motivated by these challenges, we put forward a minimal model for ultrathin superconducting films coupled to the substrate which is analytically tractable but that we expect to capture most of the relevant physics without free parameters, except the quasiparticle lifetime. However, we have found that its role is relatively minor at least in STM experiments. A refined model of the film/substrate interface, based on experimental data, would probably account for this parameter, however, this is beyond the scope of the paper.

In order to avoid the intricacies of the Kosterlitz-Thouless transition, we restrict ourselves to the low-temperature limit of weakly coupled one-band superconductor where a mean- field approach is still accurate. The film and the substrate are described by an asymmetric potential well plus a finite quasiparticle lifetime. Charge neutrality is included, although in some cases, such as in complex oxide heterostructures \cite{triscone}, it is unclear whether it applies. We note that in these materials, charge spreading across the interface alters boundary conditions at the interfaces leading to an electrostatic binding between the layers that can prevent the charge neutrality condition to hold. Disorder is not considered as the experiments can be carried out in the limit where the effect of impurities is negligible.

We report results for the superconducting gap ($\Delta$) at zero temperature as a function of the film thickness for a broad range of the parameters that define the substrate and also for different electron-phonon coupling constants. The dependence of the results on the validity of the charge neutrality condition \cite{Yu1976} is also investigated in detail. On average, the superconducting gap decreases with thickness. However, remnants of the shape resonances are still observed in some range of parameters. For a weak coupling to the substrate, and a weak electron-phonon coupling, a modest enhancement of superconductivity is observed for certain thicknesses even if the charge neutrality condition holds. Much larger enhancement is expected for material in which the charge neutrality condition does not hold. Finally, we show that this theoretical model provides a fair qualitative description of the Pb ultrathin-film experiments mentioned above.

The paper is organized as follows. In the next section, we introduce the microscopic model that describes superconductivity and the asymmetric potential well that, together with the finite lifetime, models the substrate. The model is then solved in Sec. \ref{sec:results} by a combination of mean-field and semiclassical techniques. Then, we present results of the superconducting gap as a function of the thickness for different values of the parameters. Based on this information, we discuss the range of realistic experimental settings for which it is feasible to observe shape resonance and/or an enhancement of superconductivity and discuss the relevance of these results for recent Pb ultrathin-film experiments.

\section{Model}\label{sec:model}
We put forward a model for a superconducting thin film coupled to the substrate. Superconductivity is described by a mean-field approach. The substrate is modeled by an asymmet- ric finite well that depends on the difference between the bulk chemical potential of the materials in the film and the substrate. This confinement leads to the quantization of the momentum component perpendicular to the film plane. We also introduce a finite quasiparticle finite lifetime to describe tunneling into the substrate and any other source of decoherence. Charge neutrality is also taken into account to model the interface, but we also present present results without it as we believe that in some materials it might not fully apply. We start with a description of the theoretical model employed to describe superconductivity.

\subsection{Mean field approach to superconductivity in thin films}\label{sec:model_gap_no_tau}
In a finite-size system, the BCS Hamiltonian in terms of a set of good quantum numbers is given by\begin{equation}\label{hamiltonian2}
H=\sum_{n,\sigma } \xi_{n }c^{\dagger }_{n\sigma}c_{n\sigma} - \rho \mathcal{V} \tilde \delta \sum_{n,n'}c^{\dagger}_{n\uparrow}c^{\dagger}_{n\downarrow}\tilde V_{n,n'}c_{n'\downarrow}c_{n'\uparrow},
\end{equation}
where $\rho$ is the dimensionless coupling constant, $\mathcal{V}$ is the system volume, $\tilde\delta$ is the mean level spacing [inverse of the spectral density of states at the Fermi energy ($E_F$)], $\sigma$ is the spin index, $\xi_{n}=\epsilon_{n}-\mu$, $c_{n\sigma}\ \mbox{and } c_{n\sigma}^\dagger$ are the usual quasiparticle annihilation and creation operators. The interaction matrix elements are $\tilde V_{n,n'}= \int_{\mathcal{V}} |\Psi_n(\vec r)|^2 |\Psi_{n'}(\vec r)|^2 d^3 \vec r$ where $\Psi_n(\vec r) \propto e^{i(k_yy+k_zz)}\psi_n(x)$ are the three-dimensional quasiparticle eigenfunctions with $\psi_n(x)$ the eigenstates of the one-dimensional problem in the direction perpendicular to the film.

A mean field approach to the Hamiltonian above leads to the  following Bardeen-Cooper-Schrieber (BCS) gap equation at zero temperature,
\begin{equation}\label{gap}
\Delta_n=\rho \mathcal{V} \tilde \delta \sum_{n'} \frac{\Delta_{n'}\tilde V_{n,n'}}{2\sqrt{(E_{n'}-\mu) ^2+\Delta_{n'}^2}}.
\end{equation}
The sum is restricted to those states such that $E_{n'}$ is inside the Debye window: $|E_{n'}-\mu|<\hbar \omega_D$ where $\omega_D$ is the Debye frequency.

We consider a thin film of lateral size much larger than its thickness. Therefore, the sum in Eq. (\ref{gap}) can be substituted by an integral in the in-plane momentum components, where we imposed periodic boundary conditions, and a finite sum in the perpendicular dimension. 

With the previous considerations Eq. (\ref{gap}) leads to the following system of equations for $\Delta_{k_n}$, $n\in\mathbb{N}$:
\begin{equation}\label{3rd_gap_eq}
\begin{split}
\Delta_{k_n}=\rho  \mathcal{V} \tilde\delta \frac{g_{2D}}{L^2}\sum_{n'=1}^\nu\Delta_{k_{n'}}V_{k_n,k_{n'}}\text{asinh}\left(\frac{\hbar\omega_D}{\Delta_{k_{n'}}}\right),
\end{split}
\end{equation}
where $L^2 \to \infty$ is the thin-film area,  and $V_{k_n,k_{n'}}=\int_0^adx|\psi_{k_n}(x)|^2|\psi_{k_{n'}}(x)|^2$ ($a$ is the film thickness) is obtained after having performed the $y$ and $z$ integrals in $\tilde V_{k_n,k_{n'}}$. $g_{\text{2D}}=m_{yz}L^2/(\pi\hbar^2)$ is the two-dimensional density of states and $m_{yz}$ the in-plane effective mass. The factor asinh$(\hbar\omega_D/\Delta_{k_{n'}})$ comes from the integration in the in-plane momentum components.

Since $V_{k_n,k_{n'}}$ depends on $k_n$, Eq. (\ref{3rd_gap_eq}) is a system of 
non-linear equations which leads to a momentum-dependent order parameter, 
$\Delta_{k_n}$. Assuming that the mean level spacing is much smaller than the 
bulk gap, we define the superconducting gap as \cite{degennes} the minimum 
energy needed to excite quasiparticles, namely $\min_{n}\Delta_{k_n}$. This 
observable, which is measured by STM and other spectroscopic techniques, is the 
one that we use to characterize superconductivity in the system. In order to eliminate the momentum dependence of the gap, and further simplify the calculation,  we replace  $k_n$ by $k_\nu$, the highest occupied state. In this way, an approximate solution of Eq. (\ref{3rd_gap_eq}) is simply,
\begin{equation}\label{3rd_gap_discrete}
\Delta=\frac{\hbar\omega_D}{\text{sinh}\left[K/\sum_{n=1}^{\nu} V_{k_\nu,k_n}\right]},\ \ K=\frac{\pi\hbar^2}{m_{yz}\rho \mathcal{V}\tilde \delta}.
\end{equation}
In Figs. \ref{gapPb} and \ref{gapPbsmallVs}, we show explicitly that, especially for larger values of the electron-phonon coupling constant, this is a good approximation, namely,  $\Delta \approx \min_{n}\Delta_{k_n}$. Another reason to use this additional approximation is that the corrections of the superconducting gap induced by a finite quasiparticle lifetime,  studied in Sec. \ref{sec:model_lifetime}, can easily be computed from Eq. (\ref{3rd_gap_discrete}), while a calculation from Eq. (\ref{3rd_gap_eq}) is technically very demanding. 
\subsection{Model of the thin-film coupling to the substrate}\label{sec:model_potential}
The model of the coupling between the thin film and the substrate/vacuum has three ingredients: the effective potential felt by the quasiparticles due to the substrate, the finite quasiparticle lifetime, and the charge neutrality condition. 
\vspace{-0.5cm}
\subsubsection{\bf \emph{Effective potential: An asymmetric finite well}}\label{sec:eff_pot}
The model of an asymmetric finite well has been previously implemented\cite{Czoschke2005} to study the energy spectrum and size effects in nonsuperconducting thin films. We employ the same effective potential felt by the quasiparticles in the thin film as a consequence of the substrate. As is sketched in Fig. \ref{potential}, the potential has three parameters: the  height $V_s$ of the film/substrate interface, the height $V_0$  of the film/vacuum interface, and the film thickness $a$.
\begin{figure}[H]
\begin{center}
\includegraphics[scale=0.6]{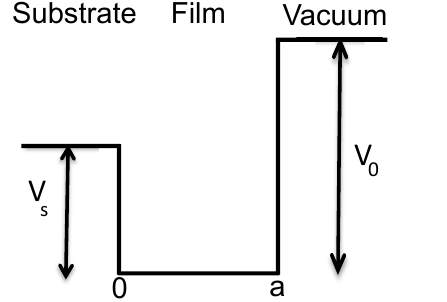}
\vspace{-3mm}
\caption{Asymmetric finite well. The values of $V_s$ and $V_0$ are discussed in Sec. \ref{sec:param}.}
\label{potential}
\end{center}
\end{figure}
\vspace{-0.6cm}
For $V_0$ we take the sum of the ionization level plus the (bulk) Fermi energy of the film material. For $V_s$ we choose the mismatch between $E_F$ and the Fermi energy of the substrate, or conduction band edge (CBE), plus an extra contribution due to the height of a Schottky barrier at the interface. In principle a more complicated potential above the CBE might give better quantitative results. However we stick to a simpler more general approach as a truly realistic potential could result in a time-dependent problem.\cite{Dijk2002} Moreover the exact details of the potential are expected to be sensitive to the substrate material. 

Before turning our attention to the solution of the Schr\"{o}dinger equation in this potential we briefly comment on the dispersion relation and the boundary condition that we have employed.\\
{\it Dispersion relation.} Following previous works\cite{Thompson1963} we use a quadratic dispersion relation but with three parameters,
\begin{equation}\label{disp_relation}
E(k)=a_0+\frac{\hbar^2}{2m_x}(k+k_L)^2,
\end{equation}
where $a_0$ and $k_L$ determine the position of the band and the effective mass $m_x$ controls the curvature. The motivation for introducing $k_L$ is that it allows to describe a back-folded conduction band (see Fig. \ref{Pb_bands}) in {\it sp} metals such as Pb and Al commonly employed in thin-film experiments \cite{Guo2004,Qin2009,Kirchmann2010,Zhang2010,Aballe2001prl}. 
\begin{figure}[t]
  \begin{center}
    \includegraphics[scale=0.35]{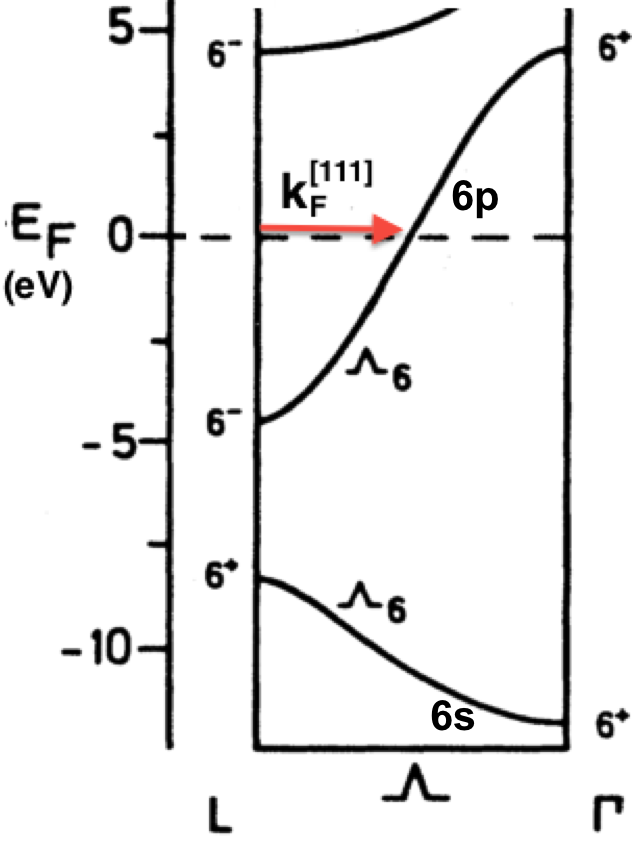}
  \end{center}
  \vspace{-0.6cm}
     \caption{(Color online) Pb band diagram in the $[111]$ direction \cite{Horn1984} .}
  \label{Pb_bands}
\end{figure}
For comparison with experimental results we take that quantization in the momentum-space direction $\Gamma L$, where $\Gamma$ and $L$ are the crystallographic points corresponding to zero momentum and $k\propto(1,1,1)$. This fixes $k_L=\pi/d$ where $d$ is the distance between atomic planes in the [111] direction. For a face-centered cubic cell $d=\frac{\sqrt{3}}{3}\times$(lattice constant). We do not consider the decrease in the lattice constant at low temperatures. On the other hand $k_F^{[111]}$ (the maximum value of $k$ in Eq. (\ref{disp_relation})) corresponds to the momentum at which the band reaches the Fermi energy. It is such that the Fermi momentum obtained from de Haas-van Alphen experiments equals $k_F^{[111]}+k_L$.

{\it BenDaniel-Duke boundary conditions.} As usual, we impose continuity of the wavefunction in both interfaces. For the continuity of the first derivative we consider the effective masses in the film and substrate. These are known as the BenDaniel-Duke boundary conditions, commonly used in heterostructures \cite{BenDaniel1966},
\begin{equation}\label{BenDaniel}
\frac{1}{m_x}\left.\frac{\partial \psi}{\partial x}\right|_{x=0}=\frac{1}{m_s}\left.\frac{\partial \psi}{\partial x}\right|_{x=0},\ \ \frac{1}{m_x}\left.\frac{\partial \psi}{\partial x}\right|_{x=a}=\frac{1}{m_e}\left.\frac{\partial \psi}{\partial x}\right|_{x=a}.
\end{equation}
We have placed the film/substrate interface at $x=0$ and the film/vacuum interface at $x=a$. We have defined $m_x$ and $m_s$ as the effective masses in the film and substrate, respectively. In the vacuum region, we have taken the free electron mass, $m_e$. 

With the previous considerations, the quantization condition for $k$ (the component perpendicular to the film) is
\begin{equation}\label{3rd_quant_cond}
\begin{split}
&(k_n+k_L)a=n\pi+\text{atan}\left(\frac{\tilde \kappa_0}{k_n+k_L}\right)+\text{atan}\left(\frac{\tilde \kappa_s}{k_n+k_L}\right),\end{split}
\end{equation}
with $n\in\mathbb{N}$, $\kappa_0=\sqrt{\frac{2m_e}{\hbar^2}[V_0-E(k_n)]}$, $\kappa_s=\sqrt{\frac{2m_e}{\hbar^2}[V_s-E(k_n)]}$, $\tilde \kappa_s=\frac{m_x}{m_s}\kappa_s$, and  $\tilde \kappa_0=\frac{m_x}{m_e}\kappa_0$. The total thin-film eigenstates are then given by
\begin{equation}\label{3rd_wavefunction}
\begin{split}
&\Psi_{\vec k}(\vec r) \propto \frac{e^{i(k_y y+k_z z)}}{L^2}\psi_{k_n}(x),\\
&\psi_{k_n}(x)=A(k_n)\sin\left(k_n x+\theta \right),\ \theta=\text{atan}\frac{k_n+k_L}{\tilde\kappa_s}\\
&A(k_n)=\left[\frac{2k_n a+\sin(2\theta )-\sin(2k_n a+2\theta )}{4k_n}+\right.\\
&\hspace{1.1cm}\left. +\frac{\sin^2\theta}{2\kappa_s}+\frac{\sin^2(k_n a+\theta )}{2\kappa_0 }\right]^{-\frac{1}{2}},
\end{split}
\end{equation}
where $L^2 \to \infty$ and $k_z$ and $k_y$, the in-plane momentum components, are subject to periodic boundary conditions. 
\subsubsection{\bf \emph{Charge neutrality}}\label{sec:model_CN}
As was mentioned earlier, Dirichlet/rigid boundary conditions at the interfaces, a key ingredient for the observation of large shape resonances, are not consistent \cite{Allen1975} with the principle of charge neutrality in film surfaces. Despite the fact that our boundary conditions allow the eigenstates to extend beyond the interface, with a typical size controlled by the step heights $V_0$ and $V_s$, charge neutrality is not yet satisfied.

In order to comply with this condition, it was proposed \cite{Sugiyama1960} to extend the potential a distance $b$ which is chosen so that surface charge neutrality holds. This shift $b$ induces a phase shift, $k b$, which together with $\theta(k)$, the phase shift induced by the potential, must satisfy,
\begin{equation}
\int_0^{k_F}[\theta(k)+kb]\ k\ \text{d}k=\frac{\pi k_F^2}{8}.
\end{equation}
 The length $b$ is obtained by using Eq. (\ref{3rd_wavefunction}) and taking into account that the quantized component of the momentum is $k_L+k_n$,
\begin{equation}\label{CN_condition}
\begin{split}
&\frac{\pi(k_L+k_F)^3}{8}=\\
&=\int_0^{k_F}dk(k+k_L)\left[(k+k_L)b+\text{atan}\left(\frac{k+k_L}{\tilde \kappa_s}\right)\right].
\end{split}
\end{equation}
As was shown elsewhere \cite{Yu1976}, the larger $b$ the stronger the average suppression of superconductivity.
Once $b$ is known, the quantized energy levels and eigenstates are computed for a well of thickness $\tilde a=a+2b$ where $a$ is the geometrical film thickness. We shall see that the charge neutrality condition also modifies the chemical potential, the matrix elements $V_{k_n,k_{n'}}$, and therefore the superconducting energy gap.  

Effectively, a finite $b$ caused by charge neutrality, amounts to a modification of the boundary conditions. Therefore, it should not change the electron density or the phonon-mediated interaction. 
We also stress that this approximate method to satisfy charge neutrality is only valid as long as $k_F^{-1}\ll a$ \cite{Garcia-Moliner}. Such condition might not be satisfied for films of only a few monolayers (ML) thick. Furthermore, in the film/substrate interface, there is a transition layer ({\it wetting layer}) \cite{Chan2003} in which the film atoms are bonded to both the substrate and other atoms of the film. Thus, it is not clear to what extent charge neutrality is applicable in this interface. Moreover, as was mentioned previously, in complex oxide heterostructures \cite{triscone} and other materials, net electric fields in the surface could severely suppress charge neutrality. 
\subsubsection{\bf \emph{Finite lifetime}}\label{sec:model_lifetime}
In this section, we introduce the last ingredient of our model for the coupling of the thin film to the substrate: a finite quasiparticle lifetime. The introduction of a finite quasiparticle lifetime is motivated by the existence of a non-zero probability of tunneling into the substrate. It is also an effective way to account for the realistic potential at the interface and other sources of quasiparticle decoherence, such as inelastic scattering. We shall see that it also plays an important role in the calculation of the superconducting gap and the chemical potential. We start with a theoretical description of the level broadening caused by a finite quasiparticle lifetime $\tau$.\\ 
{\it Smoothing of the spectral density.} From the quantization condition (\ref{3rd_quant_cond}), $n$ can be expressed as a function of the energy, $n=n(E)$. After using the Poisson summation formula, the density of states of one-dimensional quantum well is expressed as,\cite{Brack} 
\begin{equation}\label{DoS_expression}
g(E)=\frac{dn(E)}{dE}\left[1+2\sum_{l=1}^\infty \kappa(l) \cos\left(2 l\pi n(E) \right)\right]+\frac{1}{2}\delta(E-E_1),
\end{equation}
where $E_1$ is the lowest energy state. For no level broadening ($\tau\to\infty$) $\kappa(l)=1$ which results in a set of Dirac delta functions. However, as mentioned above, tunneling into the substrate or any  decoherence mechanism induces a broadening of the energy levels which effectively is described by introducing the cutoff function $\kappa(l)$. 
The precise form of $\kappa(l)$ depends to some extent of the physical mechanism that induces the broadening  but, in most cases, $\lim_{l \to \infty} \kappa(l) = 0$ at least exponentially fast. Here, following the results of Sec. $5.5$ in Ref.\cite{Brack} for the case of tunneling, we employ a Gaussian cutoff, 
\begin{equation}\label{weight_factor}
\kappa(l)\approx e^{-(lt/\tau)^2},
\end{equation}
where $t=\frac{2m_x a}{\hbar (k+k_L)}$, $m_x$ is the effective mass in the direction perpendicular to the film, and $\tau$ is the lifetime. 
Once the energy spectrum is smoothed by a finite lifetime, it is straightforward to calculate the chemical potential and the superconducting order parameter. However, before doing so, we have to evaluate the modification of the matrix elements which also enter in the gap equation.\\ 
{\it Matrix elements for states with a finite lifetime.} In order to calculate how the matrix elements are modified for a finite quasiparticle lifetime we use the approach put forward by Dijk and Nogami\cite{Dijk2002} based on the calculation of the probability of an initial state to stay inside the well.

The study of unstable or unbound eigenstates in a quantum system is an intrinsically time-dependent problem. Even though we are not interested in a time-evolution analysis, this framework allows to obtain the superposition between the initial wavefunction and the bound states. This superposition is given by the probability to stay in the film, 
 \begin{equation}
 P(t)=\int_0^L|\psi(x,t)|^2dx,
 \end{equation}
 where $\psi(x,t)$ is the initial wavefunction expressed as a linear combination of the bound and quasi-bound eigenstates. The latter can be casted as Moshinsky functions\cite{Dijk2002} which eventually escape from the potential. Therefore, for large times, $P(t)$ is given by the product of the amplitude of the bound states inside the potential, $\int_0^L dx|\psi_b(x)|^2$, multiplied by the superposition of the initial state and the bound eigenstate, namely $|c_b|^2$, where $c_b=\int_{-\infty}^\infty \psi_b(x)\psi(x,0)dx$, i.e., 
 \begin{equation}
 P(t\to\infty)\to|c_b|^2\int_0^L dx|\psi_b(x)|^2,
 \end{equation}
where $\psi_b(x)$ is a bound state of the potential well. For large times it is expected that $\psi(x,0)\to c_b\psi_b(x)$. Therefore the probability of finding the particle confined in the well will be very small provided that $c_b$ is small. It is then natural to express the matrix elements that enter in the gap equation as, 
\begin{equation}\label{3rd_mat_elem_tau}
V_{k_n,k_{n'}}=\int_0^adx|c_{b}(k_n)\psi_{k_n}(x)|^2|c_{b}(k_{n'})\psi_{k_{n'}}(x)|^2.
\end{equation}
We now rewrite the eigenstates in Eq. (\ref{3rd_wavefunction}) as
\begin{equation}\label{3rd_bounded}
\begin{split}
\psi_{k_n}^b(x)&=\begin{cases}
C_4e^{\kappa_sx},& x<0\\
C_2e^{i (k_n+k_L) x}+C_3e^{-i (k_n+k_L) x},& 0<x<a \\
C_1e^{-\kappa_0 (x-a)},& x>a
\end{cases}
\end{split}
\end{equation}
where, $\kappa_0$, $\kappa_s$ were defined previously and $C_2=\frac{C_4}{2}\left[1+\frac{\tilde\kappa_s}{i (k_n+k_L)}\right]$, $C_3=\frac{C_4}{2}\left[1-\frac{\tilde\kappa_s}{i (k_n+k_L)}\right]$, $C_1=C_2e^{i (k_n+k_L) a}+C_3e^{-i (k_n+k_L) a}$ and, from the normalization condition,
\begin{equation*}
|C_4|^{-2}=\frac{1}{2\kappa_s}+\frac{|C_1|^2}{2\kappa_0|C_4|^2}+\frac{1}{|C_4|^2}\int_0^adx|\psi_n^b(x)|^2.
\end{equation*}
We also assume that the `initial" unstable state has an energy: $E=E_n+i\Gamma/2=E_n+i\hbar/(2\tau)$, where $E_n$ is the quantized energy given by Eqs. (\ref{disp_relation}) and (\ref{3rd_quant_cond}). The initial state is given by the same type of wavefunction as Eq. (\ref{3rd_bounded}) but with the following modifications:
\begin{enumerate}[label={(\arabic*)},leftmargin=0mm,itemindent=8mm]
\item We replace $\kappa_0$ and $\kappa_s$ (see Sec. \ref{sec:eff_pot}) by $\kappa_0=\Re\left[\sqrt{\frac{2m_e}{\hbar^2}(V_0-E_n-i \frac{\hbar}{2\tau})}\right]$ and $\kappa_s=\Re\left[\sqrt{\frac{2m_e}{\hbar^2}(V_s-E_n-i \frac{\hbar}{2\tau})}\right]$.  A complex part in $\kappa_s$ or $\kappa_0$  leads to divergent terms in the matrix elements.
\item For $0<x<a$ we substitute the quantized momentum $k_n\in\mathbb{R}$ by a complex-valued $\lambda_n$. We let $\lambda_r=\Re (\lambda_n)$ and $\lambda_i=\Im(\lambda_n)$ and substitute $\lambda_n=\lambda_r+i\lambda_i$ in the dispersion relation of Eq. (\ref{disp_relation}),
with $A=\frac{2m_x}{\hbar^2}(E_n-a_0)$ and $B=\frac{2m_x}{\hbar^2}\frac{\hbar}{2\tau}$. Moreover
 $C_2$, $C_3$ and $C_1$ above are replaced by, $D_2=\frac{C_4}{2}\left[1+\frac{\tilde\kappa_s}{i (\lambda_n+k_L)}\right]$, $D_3=\frac{C_4}{2}\left[1-\frac{\tilde\kappa_s}{i (\lambda_n+k_L)}\right]$ and $D_1=D_2e^{i (\lambda_n+k_L) a}+D_3e^{-i (\lambda_n+k_L) a}$. That results in the following expression for the energy levels,
\end{enumerate}
\begin{equation}
\begin{split}
&E_n+i\frac{\hbar}{2\tau}=a_0+\frac{\hbar^2}{2m_x}(\lambda_r+i\lambda_i+k_L)^2\\
&\rightarrow\begin{cases}
&\lambda_r=-k_L+\frac{1}{\sqrt{2}}\sqrt{A+\sqrt{A^2+B^2}}\ , \\
&\begin{split}\lambda_i&=\sqrt{2}\frac{A}{\sqrt{B}}\sqrt{A+\sqrt{A^2+B^2}}+\\
&-\frac{1}{\sqrt{2}B}\left(\sqrt{A+\sqrt{A^2+B^2}}\right)^3\ .\end{split}
\end{cases}
\end{split}
\end{equation}

We have now all the necessary information to compute the initial state $\psi_{\lambda_n}(x,0)$ and then the weighting factor $c_b(k_n)=\int_{-\infty}^\infty \psi_{k_n}^b(x)\psi_{\lambda_n}(x,0)dx$. We find it more convenient to express  $c_b(k_n)$ as a function of energy $E$ since the BCS gap equation will be expressed also in terms of this variable. To that end, we substitute $k_{n'}$ in Eq. (\ref{3rd_mat_elem_tau}) by $k(E)=-k_L+\sqrt{(E-a_0)2m_x/\hbar^2}$. The resulting final expression for the matrix elements is therefore,
\begin{equation}\label{mat_elem_comp}
\begin{split}
&V(\tilde E,E)=\int_0^adx|c_b{\scriptstyle(\tilde E)}\psi_{k{\scriptstyle (\tilde E)}}^b(x)|^2|c_b{\scriptstyle(E)} \psi_{k(E)}^b(x)|^2,\\
&\hspace{1cm}c_b{\scriptstyle(E)}=\int_{-\infty}^\infty \psi_{k{\scriptstyle(E)}}^b(x)\psi_{\lambda{\scriptstyle(E)}}(x,0)dx .
\end{split} 
\end{equation}
\subsubsection{\bf \emph{Superconductivity in thin films in the presence of a substrate and a finite quasiparticle lifetime}}\label{sec:gap_finite_lifetime}
Having obtained explicit expressions for the matrix elements  
(\ref{mat_elem_comp}) and the spectral density  
(\ref{DoS_expression}), it is straightforward to find the chemical 
potential $\mu$ and the superconducting gap $\Delta$. For instance, for $\mu$,
\begin{equation}\label{3rd_mu}
\begin{split}
&N=\int_0^\mu dE_x\sum_{n'=1}^\nu\delta(E_x-E_n)\int_0^{\mu-E_x}dE_{yz}g_{2D}\\
&\rightarrow\frac{N}{V}\frac{\pi\hbar^2a}{m_{yz}}=\sum_{n'=1}^\nu(\mu-E_n)=\nu \mu -\sum_{n'=1}^\nu E_n,
\end{split}
\end{equation}
where $N/V$ is the electron density, $E_x$ and $E_{yz}$ are the energies corresponding to the out-of-plane and in-plane momentum components, respectively. The former is quantized, $E_x=E_n$, and $\nu$ is the number of occupied states. The smoothed spectrum is taken into account by replacing the sum in $n'$ by an integral in energy,
\begin{equation}\label{eq:mu_lifetime}
\sum_{n=1}^\nu(\mu-E_n)\rightarrow\int_{E_1}^\mu dE(\mu-E)g(E)=\frac{N}{V}\frac{\pi\hbar^2a}{m_{yz}}
\end{equation}
valid for $E_1<\mu<V_s$. Similarly, for the energy-dependent order parameter [Eq. (\ref{3rd_gap_eq})],
\begin{equation}\label{3rd_gap_integral_eq}
\Delta(\tilde E)=\frac{1}{K}\int_{E_1}^\mu dEg(E)\Delta(E)\text{asinh}\left(\frac{\hbar\omega_D}{\Delta(E)}\right)V(\tilde E,E),
\end{equation}
where $V(\tilde E,E)$ is given in Eq. (\ref{mat_elem_comp}) and  $K=\frac{\pi\hbar^2}{m_{yz}\rho \mathcal{V}\tilde \delta}$.
This is a non-linear Fredholm integral equation of the second kind with a non-degenerate kernel. A more tractable expression is obtained by substituting $\tilde E$ by $E_\nu$ in the previous equation. In other words the gap is approximated by the order parameter evaluated at the energy of the highest occupied state $\Delta(E_\nu)$ and, for consistency, the interaction $V(\tilde E,E)$ is replaced with $V(E_\nu,E)$.  These approximations, that neglect the energy dependence of the order parameter, result in the following algebraic expression for the energy gap,
\begin{equation}\label{3rd_gap_easier}
\Delta=\frac{\hbar\omega_D}{\text{sinh}\left[K/\int_{E_1}^\mu dE\ V(E_\nu,E)g(E)\right]},\ \ K=\frac{\pi\hbar^2}{m_{yz}\rho \mathcal{V}\tilde \delta}.
\end{equation}

Numerical results, depicted in  Figs.  \ref{gapPb} and \ref{gapPbsmallVs}, show that the substitution $\tilde E$ by $E_\nu$ or equivalently $k_n$ by $k_\nu$ in Eq. (\ref{3rd_gap_eq}) is in general a good approximation for the spectroscopic gap $\min_{n} \Delta(k_n)$, namely, $\Delta \approx \min_{n} \Delta(k_n)$. In the rest of the paper, unless it is explicitly stated otherwise, we use Eq.(\ref{3rd_gap_easier}) to compute the superconducting gap.

\section{Results}\label{sec:results}
In this section we study the superconducting order parameter $\Delta$ [Eq.(\ref{3rd_gap_easier})] for a one-band thin film coupled to a substrate as a function of film thickness and the parameters that define the substrate and the superconducting material. Our calculation includes the charge neutrality condition which should hold in Pb and other metallic superconductors except maybe in the limit of a few ML thickness. We also present results without imposing the charge neutrality condition as it is believed that in some materials, such as complex oxide heterostructures, \cite{triscone}, might not hold. We have two main motivations for this study: to provide a qualitative description of recent experiments involving Pb ultra-thin \cite{Guo2004,Zhang2010} films and also to clarify whether, in some range of parameters, size effects in thin films can enhance the critical temperature with respect to the bulk limit. 

\begin{figure}[t]
\begin{center}
\includegraphics[scale=1.6]{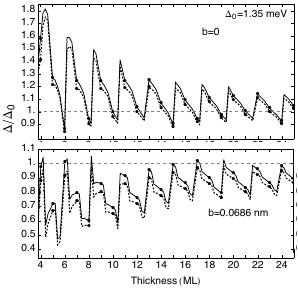}
\vspace{-0.3cm}
\caption{Superconducting order parameter $\Delta$ for a Pb thin-film coupled 
to a Si substrate. $\Delta_0$ is the bulk Pb gap. The film/interface coupling 
is modeled by an asymmetric well potential with  $V_s=E_F+1.70$ eV, 
$V_0=E_F+4.25$ eV. Level broadening is assumed to be negligible. The dimensionless coupling constant is $\rho=0.385$ and the Debye energy 
$\hbar\omega_D=9.048$ meV. The band structure parameters are given in Eq. 
(\ref{Pb_band_parameters}). The upper plot does not satisfy 
the charge neutrality condition while the lower plot does it with $b$ obtained 
from Eq.(\ref{CN_condition}). Lines show the evolution of the order parameter as a function of the thickness. Dots correspond to the estimate positions of Pb monolayers. The continuous and dashed 
lines show the difference between the $k-$dependent gap with $\Delta = \min_{n}\Delta(k_n)$ from  Eq. (\ref{3rd_gap_eq}) and the $k-$independent $\Delta$ from Eq. (\ref{3rd_gap_discrete})  
respectively. We note the pattern of shape resonances is more intricate than in Ref. \cite{Thompson1963} as a consequence of the interplay between the finite number of states in the asymmetric well potential and the more realistic dispersion relation.}
\label{gapPb}
\end{center}
\end{figure}

As was mentioned previously, the coupling to the substrate is modeled by the asymmetric finite well depicted in Fig. \ref{potential}.
The height in the film/vacuum interface $V_s$ is taken to be the bulk Fermi energy of the film plus the work function of the corresponding material. The height $V_0$ in the film/substrate interface is chosen to be the mismatch of the Fermi energies of the thin film and substrate materials plus the height of the Schottky barrier. We assign a finite quasiparticle lifetime $\tau$ to all states, including those under the barrier. This is necessary as the exact details of the potential at the interface are not well understood. Moreover, inelastic scattering and other processes will induce level broadening even when tunneling is not relevant. 
 Based on recent experiments in Pb films \cite{Kirchmann2010}, we assume a linear dependence of $\tau =\beta+ \gamma a$. The first term on the right-hand side, with $a$ the film thickness, describes tunneling into the substrate. The constant $\beta$  accounts for other size-independent mechanisms of level broadening.

\begin{figure}[t]
\begin{center}
\includegraphics[scale=1.6]{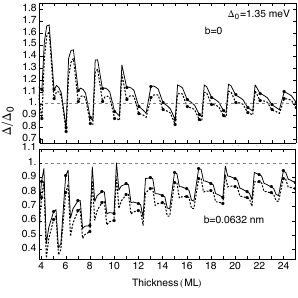}
\vspace{-0.3cm}
\caption{Superconducting order parameter $\Delta$ for a Pb film coupled to a Si substrate with no level broadening. $V_s=E_F+0.9$ eV, $V_0=E_F+4.25$ eV. $\Delta_0$ is the bulk Pb gap. The parameters correspond to those of Fig. \ref{gapPb} except for the smaller step $V_s$ and $b$ obtained from Eq. (\ref{CN_condition}). The continuous line corresponds to the $k-$dependent gap with $\Delta = \min_{n}\Delta(k_n)$ from  Eq. (\ref{3rd_gap_eq}) and the dashed line corresponds to the $k-$independent $\Delta$ from Eq. (\ref{3rd_gap_discrete}). The $k-$independent approximation becomes less accurate as the step height decreases. As was expected, reducing the height of the film/substrate interface increases the leaking of probability out of the film which, reduces both the superconducting gap $\Delta$ and the effect of the charge neutrality condition measured by $b$. Moreover, the pattern of shape resonances is richer as the number of bound states is smaller in this case.}
\label{gapPbsmallVs}
\end{center}
\end{figure}
\subsection{Parameters: Pb films grown over a Si substrate }\label{sec:param}
In this section, we introduce the range of parameters that we use in the calculation of the superconducting gap. 
First, we focus in one of the best studied settings \cite{Guo2004}: Pb thin films grown over a Si substrate.

As discussed in Sec. \ref{sec:eff_pot}, the dispersion relation is described in terms of three parameters $k_L$, $a_0$, and $m_x$, the effective mass in the direction perpendicular to the film. The first is fixed by the inter-atomic plane distance $k_L=\pi/d$ while the other two are set in order to describe the bulk Pb Fermi level and the minimum of the Pb band in the crystallographic $L$ point.
Other relevant  parameters in the calculation of the chemical potential and the energy gap are the in-plane effective mass and the electron density $\frac{N}{V}$. 
The exact value of the in-plane effective mass $m_{yz}$ and its dependence with the film thickness are still a subject of discussion \cite{Upton2004,Dil2006}. We are not interested to study this effect at the moment and fix it to a constant value. We also impose that for a very large thickness, Eq.(\ref{3rd_mu}) leads to a chemical potential equal to the Fermi energy. With these considerations in mind we now state the values of the parameters we employ,
\begin{equation}\label{Pb_band_parameters}
\begin{split}
&m_x=1.180m_e,\ \ a_0=1.57\ \mbox{eV},\\
&k_L=\frac{\pi}{d}=10.99\mbox{ nm}^{-1},\ \ k_F^{[111]}=0.450\frac{\pi}{d}=4.95\mbox{ nm}^{-1},\\ 
&m_{yz}=1.380m_e,\ \ \frac{N}{V}=20.69\mbox{ nm}^{-3},\ E_F=9.77\mbox{ eV}.
\end{split}
\end{equation}
$m_x$ is close to the value reported in the literature $m_x=1.14m_e$ \cite{Anderson1965}, while $k_F^{[111]}$ is taken from Refs. \cite{Qin2009,Pan2011}. $d=\frac{\sqrt{3}}{3}0.4951=0.2858$ nm is the distance between (111) planes. With these parameters, the energy of the band that we study (see Fig. \ref{Pb_bands}) ranges from $5.47$ eV at the $L$ point to the Fermi energy, $9.77$ eV \cite{Horn1984}. Finally, for the substrate effective mass in the direction perpendicular to the interface we take $m_s=0.28m_e$ \cite{Dresselhaus1955}.

The next step to model the thin film is to the impose the charge neutrality condition. From Eq.\ref{CN_condition}) and by using the parameters above, we have found that, in order to comply with this condition, the thin-film thickness $a$ must be effectively extended to $a \to a +2b$ with,
\begin{equation}
b=0.0686\mbox{ nm}.
\end{equation}
This is less than half the distance between (111) atomic planes approximately. This correction is smaller than for a free-standing film\cite{Yu1976} since the potential from Fig. \ref{potential}, in contrast to an infinite potential well, allows already leaking of probability out of the film.

The parameters of the asymmetric potential that characterize the substrate are chosen as follows:
for the height in the film/vacuum interface, we take the work function above the Fermi energy, $W_F=4.25$ eV. 
The height of potential at the substrate--thin-film interface is the mismatch between the CBE of the substrate and the bulk Fermi energy of the film plus the height of the Schottky barrier. For Pb/Si films, the Si CBE is $0.8$ eV above the Pb Fermi energy,\cite{Kirchmann2010} while we use $0.9$ eV for the height of the Schottky barrier corresponding to the $(\sqrt{3}\times\sqrt{3})R30^\circ$ orientation \cite{Heslinga1990}. The asymmetric well potential is therefore characterized by, 
\begin{equation}\label{Pb_steps}
\begin{split}
&V_s=E_F+0.80\mbox{eV}+0.90\mbox{eV}=11.47\mbox{ eV, }\\
&V_0=14.02\mbox{ eV}. 
\end{split}
\end{equation}

Pb is not a weakly coupled superconductor so in principle the Eliashberg 
theory of superconductivity is more suitable to describe its properties. However 
the BCS prediction for the temperature dependence of the superconducting order 
parameter describes the experimental data reasonably well 
\cite{Eom2006,Qin2009}, even for a single Pb atomic monolayer \cite{Zhang2010}. 
For that reason, and taking into account that our main interest is the 
superconducting gap, we have decided to use the simpler BCS introduced previously to describe 
size effect in this material.  We employ the following values of the Debye 
energy and the dimensionless coupling constant,\cite{Poole}
 \begin{equation}
\hbar\omega_D=9.048\mbox{ meV, }\ \rho=0.385\rightarrow \Delta_0=1.35\mbox{ meV,}
\end{equation}

The last element in our model is the quasipartcle lifetime $\tau$. For sufficiently small $\tau =\tau_{0}$, we expect suppression of all size effects. This scale corresponds to a level broadening comparable to the one-dimensional mean level spacing, $\frac{\Gamma_0}{2}\sim\delta_{\text{1D}}=\frac{1}{g_{\text{1D}}\scriptstyle{(E_F)}}=\frac{2}{a}\frac{E_F-a_0}{k_F}$, where $a_0$ is defined in Eq. (\ref{disp_relation}). The lifetime related to this energy is,
\begin{equation}\label{tau_critical}
\tau_{0}=\frac{2\hbar}{\Gamma_0}\sim\frac{\hbar k_F}{2(E_F-a_0)}a,
\end{equation}
which for Pb is $\tau_{0}=0.18(N-1)$ fs,  where $N$ is the number of monolayers. Therefore, for $\tau \gg \tau_0$, decoherence effects are small but for $\tau \sim \tau_0$ size effects related to quantum coherence will be strongly suppressed. 

We employ a simple linear model for the lifetime, 
\begin{equation}
\tau=\beta+\gamma  a.
\end{equation}
with $\beta,\gamma>0$ and $a$ the thickness.
As was explained above, if tunneling into the substrate is relevant,  $\tau$ is expected to be proportional to the thickness $a$. This a good approximation provided the tunneling probability is constant for every thickness considered. In other words, we assume the interface potential does not change as the film thickness changes.
Additionally, we include a constant term $\beta$ which accounts for  other decoherence effects. 
In principle, it is tempting to relate $\beta$ to level broadening by electron-electron scattering. The scattering rate can be estimated from Fermi liquid theory: $\Gamma_{e-e}=\alpha (E-E_F)^2$ by substituting $E-E_F$ by $\delta_{\text{1D}}$, the one-dimensional mean level spacing. This yields a scattering rate $\Gamma_{e-e}\simeq\frac{0.02}{a^2}$ eV, with $a$ in nm, which is more than two orders of magnitude smaller than the critical broadening $\Gamma_0 \propto 1/\tau_0$. Therefore, it seems that it does not play a significant role in our system. We take $\beta \sim \tau_0$ so that, by tuning $\gamma$, we can study  
the full range of corrections induced by a finite lifetime. In that way we can determine, for a given set of parameters, the range of $\tau$'s for which corrections due to a fine lifetime are relevant.
Finally we also assume that the smoothing of the spectral density is well described by Eq. (\ref{DoS_expression}).

\subsection{Size effects in the superconducting energy gap}
In this section, 
we first investigate the superconducting order parameter 
for Pb thin films coupled to a Si substrate in the absence of tunneling. We 
study the role of the coupling to the substrate in the shape resonances as well the effect of charge neutrality in 
suppressing superconductivity. We then discuss the smoothing of size effects by a finite lifetime $\tau$. 
Finally, we move from Pb in order to investigate size effects in a weakly coupled 
superconducting thin film by simply modifying the Debye energy and 
dimensionless coupling constant while leaving the rest of the parameters 
unchanged.

\subsubsection{\bf \emph{Infinite lifetime}}\label{sec:inf_tau}
In this section, we consider the limit of no level broadening ($\tau \to \infty$) with the substrate described by the asymmetric well  (Fig. \ref{potential}). The momentum-dependent order parameter is obtained from Eq. 
(\ref{3rd_gap_eq}) where,  as was mentioned in Sec. \ref{sec:model_gap_no_tau}, 
the superconducting gap is the minimum of the order parameter. We also 
approximate the solution of Eq. 
(\ref{3rd_gap_eq}) by assuming a $k-$independent order parameter  Eq. (\ref{3rd_gap_discrete}). 
In Figs. \ref{gapPb} and \ref{gapPbsmallVs}, we analyze the differences between the two predictions for different values of the asymmetric potential.

 The pattern of shape resonances is 
qualitatively similar in both cases. It is clear however that the approximate solution (\ref{3rd_gap_discrete}) is 
always below the actual gap (\ref{3rd_gap_eq}). This difference is more evident in Fig. 
\ref{gapPbsmallVs} where the potential is shallower and the coupling to the substrate is therefore stronger. Given that the system of equations (\ref{3rd_gap_eq}), can easily be 
 solved without any approximation, in principle there is no substantial advantage in using the
 approximate solution. However, once a finite lifetime is considered, the approximate solution to the order parameter is still easily obtained from Eq. 
 (\ref{3rd_gap_easier}), while the momentum-dependent order parameter ought to 
 be calculated from the integral equation (\ref{3rd_gap_integral_eq}), 
 is much more difficult to solve. For that reason, and because the results are qualitatively similar, we stick to Eq. (\ref{3rd_gap_easier}) to compute the superconducting gap in the rest of the paper.\\

 We are now ready to study the role of the substrate in our results. 
In Fig. \ref{gapPb}, we compute the superconducting gap with the parameters defined in Eq. (\ref{Pb_band_parameters}). 
For the film/substrate height we take $V_s=E_F+1.7$ eV which accounts for the 
Fermi level mismatch with the substrate plus the height of the Schottky 
barrier. In Fig. \ref{gapPbsmallVs} we remove the Schottky barrier contribution leaving 
$V_s=E_F+0.8$ eV where $0.8$ eV corresponds to the Fermi level mismatch with 
the substrate. As was expected from the model used to couple the film to the 
substrate, introduced in Sec. \ref{sec:eff_pot}, and the expressions of the BCS interaction 
matrix elements, introduced in Sec. \ref{sec:model_gap_no_tau}, we observe that a 
decrease of the potential height is accompanied by an average suppression of superconductivity. This is a simple consequence of two facts, the states are more extended into the substrate and the potential has 
less bound states.\\
Moreover, as a consequence of the coupling to the substrate, the pattern of shape resonances differs from that of an infinite well \cite{Thompson1963} where $\Delta$ decreases monotonically with the thickness until another state is occupied. Our results, depicted in Figs. \ref{gapPb} and \ref{gapPbsmallVs}, show as the potential height decreases, the momentum dependence of the order parameter becomes more relevant yielding an additional non monotonic behavior with additional features. These extra features are originated by the combined effect of the momentum-dependent interaction, the finite number of states in the asymmetric well,  and the off-centered dispersion relation (\ref{disp_relation}). The maxima and minima do not necessarily correspond to a different number of occupied states in the considered band ($6p$ band in Fig. \ref{Pb_bands}).\\
 As the thickness increases, the occupied states are lowered in the potential well (more bounded) which yields the moderate, smooth increase observed in the above plots between two prominent peaks. For some thickness the lower state in the upper band reaches the minimum at the $L$ point and thereupon this electron occupies a state in the lower  $6s$ band. At the same time another available state in the $6p$ band is occupied and thus, even though the number of occupied states in the $6p$ band is the same, these are higher in energy (less bounded)  yielding a sudden decrease. Finally, for a larger thickness the number of occupied states in the $6p$ band increases and a large increase is observed. 

The previous figures show the effect of charge neutrality is qualitatively 
similar to that in an infinite potential well \cite{Yu1976}, the average 
$\Delta$ decreases the thinner the film is. Furthermore, as $V_s$ decreases, the 
charge neutrality correction, measured by $b$, is smaller. In other words, both 
charge neutrality and a reduction of the potential height have a similar effect: to suppress superconductivity so that for all thicknesses the gap is below the bulk limit. 

As was mentioned in Sec. \ref{sec:model_CN}, the method used to impose charge 
neutrality is only valid in the limit $k_F^{-1}\ll a$. For Pb films in the range of 
thickness studied $k_F^{-1}\leq 15\times$thickness, however, the validity of the 
method is less clear as the thickness decreases. Furthermore, it is still under 
discussion as to whether, or to what extent, this condition realizes in realistic 
nano-structures.\cite{Chaib2005}
\subsubsection{\bf \emph{Finite lifetime}}
We now study the role of a finite lifetime $\tau$ that describes tunneling 
out of the film and other sources of decoherence. Following results of previous sections we use the 
smoother density of states  (\ref{DoS_expression}) to compute first the 
chemical potential  (\ref{eq:mu_lifetime}) and finally the superconducting 
energy gap (\ref{3rd_gap_easier}).

We assume a linear dependence of 
$\tau$ with the thickness. Shape resonances in the 
superconducting gap at zero temperature, depicted in Fig. \ref{gapPbtau}, are 
suppressed for $\tau$ comparable to $\tau_0$ [Eq. (\ref{tau_critical})], the time scale related to the mean level spacing in the asymmetric well potential. More 
precisely, for Pb/Si films of less than $10$ ML the suppression is considerable 
when $\tau\le 10\tau_0$ (see the blue data). This suppression is clearer if one considers the experimentally accessible thicknesses  (integer numbers of monolayers): the red and blue dots in the previous figure show that it is indeed expected to measure small oscillations in $\Delta$ of a Pb thin film.\\
For smaller $\tau$, the effect of level broadening completely smears size effects however in this range of lifetime the leading effect is to suppress superconductivity, $\Delta\rightarrow0$, due to the modification introduced in the interaction matrix elements [Eq.(\ref{mat_elem_comp})]. 

\begin{figure}[H]
\includegraphics[scale=1.4]{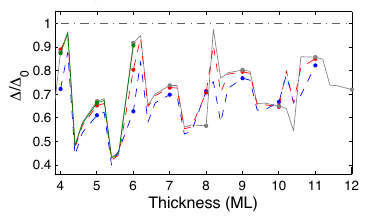}
\vspace{-0.5cm}
\caption{(Color online) Superconducting order parameter $\Delta$ at $T=0$ for Pb films on a Si substrate for different quasiparticle lifetimes. $\Delta_0$ is the bulk Pb gap, $\rho=0.385$, and $\hbar\omega_D=9.048$ meV. Results include the charge neutrality condition. The gray line corresponds to the limit of no level broadening ($\tau \to \infty$) [Eq. (\ref{3rd_gap_discrete})] (green) $\tau\mbox{(fs)}=\tau_0 +3N$,(red) $\tau\mbox{(fs)}=\tau_0+ 1.5N$, and (blue) $\tau\mbox{(fs)}=\tau_0+0.7N$ where $N$ is the number of ML. These values are chosen in order to estimate the range in which finite-$\tau$ corrections are relevant. For less than $10$ML this occurs for $\tau\le10\tau_0$, with $\tau_0$ given by Eq. (\ref{tau_critical}). Dots correspond to the exact position of the Pb monolayers.}
\label{gapPbtau}
\end{figure}

It is also 
clear that, especially for small thicknesses, charge neutrality is the dominant 
mechanism for suppression of superconductivity. It reduces substantially the 
value of the gap $\Delta$ so that it is under the bulk value in the full range of parameters 
investigated. 

\subsubsection{\bf \emph{Comparison with experiments}}
Recent STM experiments on a single monolayer of Pb deposited on Si \cite{Zhang2010} indicate sharp peaks in the tunneling data which correspond, approximately, to $\tau$, two orders of magnitude larger than the one used here. This suggests that in this setting, tunneling into the substrate is negligible even for one atomic monolayer. 

In this limit  (see results depicted in Figs. \ref{gapPb} and \ref{gapPbsmallVs}), we have observed that, in agreement with the experimental results, size effects in the presence of the substrate, and including the charge neutrality condition, lead to a superconducting gap which is below the bulk limit. As the film thickness approaches the 1-ML limit, the exponential tails of the thin-film eigenstates into the substrate become longer and therefore we expect a strong suppression of the gap. Strictly speaking, this limit can not be studied quantitatively within our model since we neglect other effects that might become relevant in this situation, such as surface phonons or the enhancement of Coulomb interactions. However, our model still predicts a strong suppression superconductivity.
 
The results presented in Fig. \ref{gapPbtau}, which include a finite lifetime, provide a good description of the superconducting gap in thin Pb/Si films obtained by transport measurements \cite{Guo2004}.  Our model reproduces correctly the small oscillations of the critical temperature observed experimentally in the region  $\sim 20$ML, the gradual suppression of the average gap as thickness is reduced and the smoothing of shape resonance for $\leq 15$ML. 

We note that the main difference between the two experiments is the presence of a capped layer in Ref. \cite{Guo2004} needed to carry out transport measurements. Even if tunneling into the substrate is negligible, as the STM results of Ref. \cite{Zhang2010} suggest, the film coupling to the overlayer still causes important decoherence effects which in our model correspond to a much smaller choice of $\tau$ than in the description of the STM experiment. 

In summary, by tuning $\tau$ we are able to describe qualitatively the experimental results of Refs. \cite{Guo2004,Zhang2010}. We note in the particular case of Pb/Si films the Si band gap in the crystallographic direction perpendicular to the interface yields a strong state confinement in the Pb film and thus tunneling into the substrate is suppressed. However, for other cases, such as Al films, the confinement is not caused \cite{Aballe2001} by a band gap and thus tunneling can be a relevant source of decoherence that can be included with the model presented in Sec. \ref{sec:model_lifetime}. Nonetheless, in this case it is likely that more sophisticated theoretical models of the interface are necessary for a quantitative description of the experimental results. 
\subsubsection{\bf \emph{Weakly coupled superconductors}}
From the results of the previous section it seems rather unrealistic, at least in Pb, to enhance superconductivity by size effects. Lead is a strong coupled superconductor so it would be interesting to explore whether size effects are stronger in materials characterized by a weaker coupling constant $\rho$. Indeed, from Eq. (\ref{3rd_gap_discrete}) it is straightforward to show that the first-order correction to $\Delta$ is inversely proportional to the coupling constant $\rho$. Therefore, the smaller $\rho$, the larger the finite size correction. Even if charge neutrality applies, the oscillations of the superconducting gap are expected to show higher maxima, with respect to the bulk limit, for smaller $\rho$ which might lead to an enhancement of superconductivity. 
In this section, the dimensionless coupling constant is decreased to $\rho=0.180$ and the Debye energy is set to $\hbar\omega=33.882$ meV. We analyze the case of $\tau\rightarrow\infty$ and maintain the same parameters for the band structure and the asymmetric potential  [Eqs. (\ref{Pb_band_parameters})-(\ref{Pb_steps})], as here our goal is to explore the dependence on the coupling constant $\rho$ rather than to model a specific material.

\begin{figure}[t]
\hspace{-3mm}\includegraphics[scale=1.7]{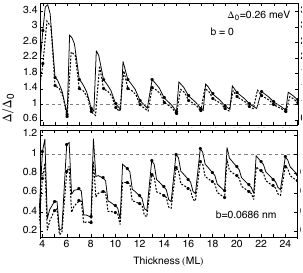}
\vspace{-0.3cm}
\caption{Superconducting gap $\Delta$ in units of the bulk gap $\Delta_0$ for  $\tau\rightarrow\infty$.  All parameters are equal to those of Fig. \ref{gapPb} except the dimensionless coupling constant $\rho=0.180$ and the Debye energy $\hbar\omega=33.882$ meV. Results in the upper plot do not satisfy the charge neutrality condition while the lower plot does include it with $b$ obtained from Eq.(\ref{CN_condition}). The gap $\Delta=\min_n\Delta(k_n)$ (continuous line) shows a moderate enhancement of superconductivity, even when charge neutrality is imposed. By contrast no enhancement is observed (dashed line) in the approximate solution Eq. (\ref{3rd_gap_discrete}).}
\label{gapXtau}
\end{figure}

The results, depicted in 
Fig. \ref{gapXtau}, show a considerable enhancement of superconductivity when charge neutrality is not imposed. If it is included a moderate enhancement is still observed for a few values of the thickness. As for Pb,  see Figs. \ref{gapPb} and \ref{gapPbsmallVs}, the exact solution $\Delta=\min_n\Delta(k_n)$ (continuous line), Eq. (\ref{3rd_gap_eq}), predicts a larger gap than the approximation, Eq. (\ref{3rd_gap_discrete}) (dashed line). Indeed we observe a net enhancement only in the case $\Delta = \min_{n} \Delta(k_n)$. This is a strong suggestion that an enhancement of superconductivity might occur for a finite lifetime provided that the gap is computed directly from Eq. (\ref{3rd_gap_integral_eq}). We note that the approximate solution (blue) shows no enhancement of $\Delta$ with respect to the bulk limit even for $\tau\rightarrow\infty$ so finite $\tau$ corrections, Eq. (\ref{3rd_gap_easier}), would induce a further suppression of the energy gap.  

In summary, weakly coupled superconducting materials are more promising candidates to observe an enhancement of superconductivity in thin films and nanostructures provided the quasiparticle lifetime is much larger than $\tau_0$, Eq. (\ref{tau_critical}). 

\section{Conclusions}
We have investigated analytically the effect of the substrate on superconducting thin films. We aim to provide a description of recent Pb thin-film experiments and also to identify a region of parameters in which size effects could enhance superconductivity. Superconductivity is modeled by a mean-field formalism. The model of the coupling of the thin film to the substrate has three ingredients: an asymmetric quantum well, a finite quasiparticle lifetime  (that describes tunneling into the substrate and other decoherence mechanisms), and the charge neutrality condition on the interfaces. For Pb on a Si substrate, we observe small oscillations, remnants of shape resonances,  of the energy gap as thickness is decreased but always below the bulk limit for realistic values of the quasiparticle lifetime and the interface potential. This is fully consistent with the transport measurements of Ref. \cite{Guo2004} in which a capped layer induces additional level broadening, for sufficiently thin films. In the limit of negligible broadening our results are also consistent with the {\it in situ} STM experiments of Ref.\cite{Zhang2010} in which a capped layer is not present. For materials with a smaller electron-phonon coupling constant, size effects are stronger. We identify a range of parameters, $\tau > 20$fs, thicknesses $\leq 10$ML, for which a modest enhancement of superconductivity is feasible even if charge neutrality holds. A stronger enhancement is expected provided that charge neutrality does not apply. This seems to be the case in complex oxide heterostructures. 
\acknowledgements
\vspace{-0.3cm}
ARB acknowledges support from a la Caixa foundation fellowship. AMG acknowledges supports from EPSRC, grant No. EP/I004637/1,  FCT, grant PTDC/FIS/111348/2009 and a Marie Curie International Reintegration Grant
PIRG07-GA-2010-268172. 
\bibliography{biblio}

\end{document}